%% file: rev2.tex
\documentclass[twocolumn,trackchanges]{aastex631}
\accepted{January 20, 2023}

\submitjournal{\apjl}

\usepackage{amsmath,environ}
\usepackage{listings}
\usepackage{float}
\usepackage{color}
\usepackage{mathtools}

\defcitealias{pettini04}{PP04}
\newcommand{\ha}{H$\alpha$\,}
\newcommand{\cii}{[C\,{\sc ii}]\,}
\newcommand{\spt}{SPT0418-47\,}

\begin{document}

\title{Discovery of a Dusty, Chemically Mature Companion to a z$\sim$4 Starburst Galaxy in JWST ERS Data}

\correspondingauthor{Bo Peng}
\email{bp392@cornell.edu}

\author[0000-0002-1605-0032]{Bo Peng}
\affiliation{Department of Astronomy, Cornell University, Ithaca, NY 14853, USA}

\author[0000-0002-4444-8929]{Amit Vishwas}
\affiliation{Cornell Center for Astrophysics and Planetary Science, Cornell University, Ithaca, NY 14853, USA}

\author{Gordon Stacey}
\affiliation{Department of Astronomy, Cornell University, Ithaca, NY 14853, USA}

\author{Thomas Nikola}
\affiliation{Cornell Center for Astrophysics and Planetary Science, Cornell University, Ithaca, NY 14853, USA}

\author[0000-0003-1874-7498]{Cody Lamarche}
\affiliation{Department of Physics, Winona State University, Winona, MN 55987, USA}

\author[0000-0002-8513-2971]{Christopher Rooney}
\affiliation{Department of Astronomy, Cornell University, Ithaca, NY 14853, USA}

\author[0000-0002-1895-0528]{Catie Ball}
\affiliation{Department of Astronomy, Cornell University, Ithaca, NY 14853, USA}

\author[0000-0001-6266-0213]{Carl Ferkinhoff}
\affiliation{Department of Physics, Winona State University, Winona, MN 55987, USA}

\author[0000-0002-8712-369X]{Henrik Spoon}
\affiliation{Cornell Center for Astrophysics and Planetary Science, Cornell University, Ithaca, NY 14853, USA}

\begin{abstract}

We report the discovery of two companion sources to a strongly lensed galaxy \spt (``ring'') at redshift 4.225, targeted by the JWST Early Release Science program. 
We confirm that these sources are at a similar redshift as the ring based on \ha detected in the NIRSpec spectrum, and \cii\,$\lambda$158 \micron\, line from ALMA. 
Using multiple spectral lines detected in JWST/NIRSpec, the rest-frame optical to infrared images from NIRCam and MIRI, and far-infrared (FIR) dust continuum detected by ALMA, we argue that the newly discovered sources are actually lensed images of the same companion galaxy, hereafter referred to as SPT0418-SE (``SE''), located within 5 kpc in the source plane of the ring. 
The star formation rate derived using \cii and dust continuum puts a lower limit of 17 M$_\sun$/yr, while the SFR$_\mathrm{H\alpha}$ is estimated to be $>$2 times lower, thereby confirming that SE is a dust obscured star-forming galaxy. 
Analysis using optical strong line diagnostics suggests that SE has near solar elemental abundance, while the ring appears to have super-solar metallicity O/H and N/O. 
We attempt to reconcile the high metallicity in this system by invoking early onset of star formation with continuous high star forming efficiency, or that optical strong line diagnostics need revision at high redshift. 
We suggest that \spt resides in a massive dark matter halo with yet to be discovered neighbors. 
This work highlights the importance of joint analysis of JWST and ALMA data for a deep and complete picture of the early Universe. 

\end{abstract}


\section{Introduction}
\label{sec:intro}

Gravitationally lensed galaxies provide us with unique opportunities to discover and study distant galaxies in detail. 
Hundreds of ultra luminous infrared galaxies (ULIRGs, $\mathrm{L_{IR}} > 10^{12} \mathrm{L_\sun}$) have been discovered in (sub)millimeter wavelengths \citep[c.f.][]{ivison98,scott08,weiss13,canameras15,harrington16,berman22}.
Later studies reveal most of the them as strongly lensed galaxies at high redshift \citep{spilker16,vieira13}. 
These dusty star forming galaxies (DSFG) are characterised by very high star formation rate (SFR) and dust mass, and they are thought to be the progenitors of massive early type galaxies seen in the local Universe \citep{casey14}.
The highest SFR systems likely trace the most massive halos in the Universe, and some of these galaxies are later identified as galaxy pairs or proto-clusters \citep{capak11,aguirre13,marrone18,miller18}.
Studying the stellar population and physical conditions in these galaxies can help us understand how such extremely dusty and massive systems have formed and evolved when the Universe was less than 2 billion years old.

SPT-S J041839-4751.8 (hereafter \spt) is a DSFG discovered in SPT-SZ survey at redshift 4.2248 \citep{weiss13}. 
Studies using dust continuum and various molecular and fine-structure lines found that \spt has very high lensing magnification $\mu = 32.3 \pm 2.5$ at 836 $\mu$m (rest-frame 160 $\mu$m), and high intrinsic SFR $\sim$ 300 M$_\sun$/yr with a moderate stellar mass $\sim 1.2 \times 10^{10}\ \mathrm{M_\sun}$ \citep{aravena16,bothwell17,debreuck19,rizzo20}. 
It is also included in the JWST Early Release Science program, which for the first time, enables us to study its physical properties in hot ionized gas through optical emission lines, as well as stellar population and distribution. 
In this paper, we report a serendipitous discovery of a dusty star-forming companion galaxy of \spt in JWST/NIRSpec data.

\section{Observational Data}
\label{sec:observations}

\subsection{JWST Data}
\label{sec:jwst}

We made use of the JWST dataset targeting \spt by the Early Release Science Program TEMPLATES (Targeting Extremely Magnified Panchromatic Lensed Arcs and Their Extended Star formation, ID: 1355, PI: Jane Rigby). 
We downloaded the uncalibrated data from the MAST archive\footnote{The data and association files are accessible via\dataset[10.17909/vr1e-rm03]{https://doi.org/10.17909/vr1e-rm03}.}, and reduced the data with the JWST pipeline version 1.8.5 and the calibration reference file context jwst\_1027.pmap.

The dataset consists of a NIRSpec IFU spectral cube, as well as NIRCam and MIRI images.
The NIRSpec IFU observation were carried out using the F290LP filter, with wavelength coverage 2.87-5.27\,\micron. 
The target was observed with the NIRCam instrument in six filters (F115W, F150W, F200W, F277W, F356W, F444W) and the MIRI imager in seven filters (F560W, F770W, F1000W, F1280W, F1500W, F1800W, F2100W), providing near continuous photometry from rest-frame $\lambda_\mathrm{rest}=$\,0.22\,\micron\ to 4\,\micron.

For positional alignment, we shift both the images and the spectral cube by matching the center of the foreground lensing galaxy model with the coordinate of the galaxy in the Gaia catalog \citep{gaia,gaiadr3}.
This alignment method make use of the bright nuclear feature in the foreground elliptical galaxy, and the local shifting solution avoids the effect of image distortion on the detector.
We applied a minor astrometric correction of $\leq$0\farcs2 to the NIRCam images and up to 0\farcs5 in NIRSpec IFU data as well as MIRI images obtained from archive. 
We also manually flagged bad pixels and resampled the image coordinate grid to facilitate comparison between different bands and instruments.

Because this is a galaxy-galaxy lens system and the foreground elliptical galaxy, functioning as the lens for the high-z background galaxy, contaminates the lensed Einstein ring strongly in the near infrared bands. 
We have taken several steps in data processing to mitigate the contamination from the foreground lensing galaxy emission and produce lens-subtracted spectral cube and images revealing the full extent of the emission from the background lensed galaxies.
The pixels at the centre of the lensing galaxy show calibration artifacts in NIRSpec data, and the core has a bright point source.
So, we flagged the pixels within a radius of 0\farcs25 from the center of the lensing galaxy in the first step.
Because the emission of the lensing galaxy strongly contaminates the the high-z source, we further fit a model including an extended elliptical S\'ersic profil and a compact 2-D Gaussian profile, both convolved with the JWST point spread function (PSF)\footnote{The PSF is computed by using WebbPSF version 1.1.1}, for the lensing galaxy and subtract it from the data.
This procedure removes most of the emission of the lensing galaxy, but the subtraction is not perfect and adds to the uncertainty in the vicinity of the lensing galaxy nucleus, especially at short wavelength $\lambda<5$\,\micron.

\subsection{ALMA}
\label{sec:alma}

We use the deep ALMA observations in the data archive (ID:2016.1.01374.S, PI: Hezaveh), which targets \spt\ to study the presence of dark matter substructure through strong lensing measurements. 
We use the execution block that covers the \cii fine-structure line ($^{3}$P$_{3/2}\rightarrow^{3}$P$_{1/2}$, $\lambda_\mathrm{rest}$=157.741\,\micron, $\lambda_\mathrm{obs}$=824.165\,\micron) and the underlying dust continuum.

The source was observed with ALMA on 2016 October 25, using 43 12\,m antennas with baselines ranging from 18.6\,m to 1.4\,km and PWV between 0.65-0.75\,mm. 
Two spectral windows in the upper sideband of the receiver cover the center frequency of the \cii line ($\nu_{obs}$=363.794\,GHz). 
The spectral windows in the lower sideband cover a bandwidth of 3.75\,GHz around the continuum at rest-frame 160 \micron ($\mu_{obs}\sim$835\,\micron).
J0455-4615 and J0439-4522 were observed for amplitude and phase calibration.
J0538-4405 was used as the bandpass calibrator and J0519-4546 as the flux calibrator. 
The integration time for our science target is 32.5 minutes.

The data is retrieved from the ALMA data archive and reduced by manually running the pipeline in the Common Astronomy and Software Application (CASA) version 4.7.2 \citep{casa}.
In particular, we take care to avoid automatic flagging of line channels that are at the edge of the spectral windows. 
We created a continuum map by collapsing the line-free channels covering a bandwidth of 5.47\,GHz and using natural weighting. 
The image is also corrected for the primary beam as the field center specified for the ALMA observations was offset by 4\farcs9 from the location of the target.
The resulting image has a beam size of 0\farcs18$\times$0\farcs17 with a position angle of -87.6$\degr$ and achieved a 1-$\sigma$ sensitivity of 56\,$\mu$Jy\,beam$^{-1}$. 
We also imaged the line channels to create a spectral cube with a 50 km/s resolution, reaching 1-$\sigma$ noise $\sim$0.6\,mJy\,beam$^{-1}$\,ch$^{-1}$.

\section{Results}
\label{sec:results}

\subsection{Discovery}
\label{sec:discovery}

\begin{figure*}
	\centering
	\includegraphics[width=1\textwidth]{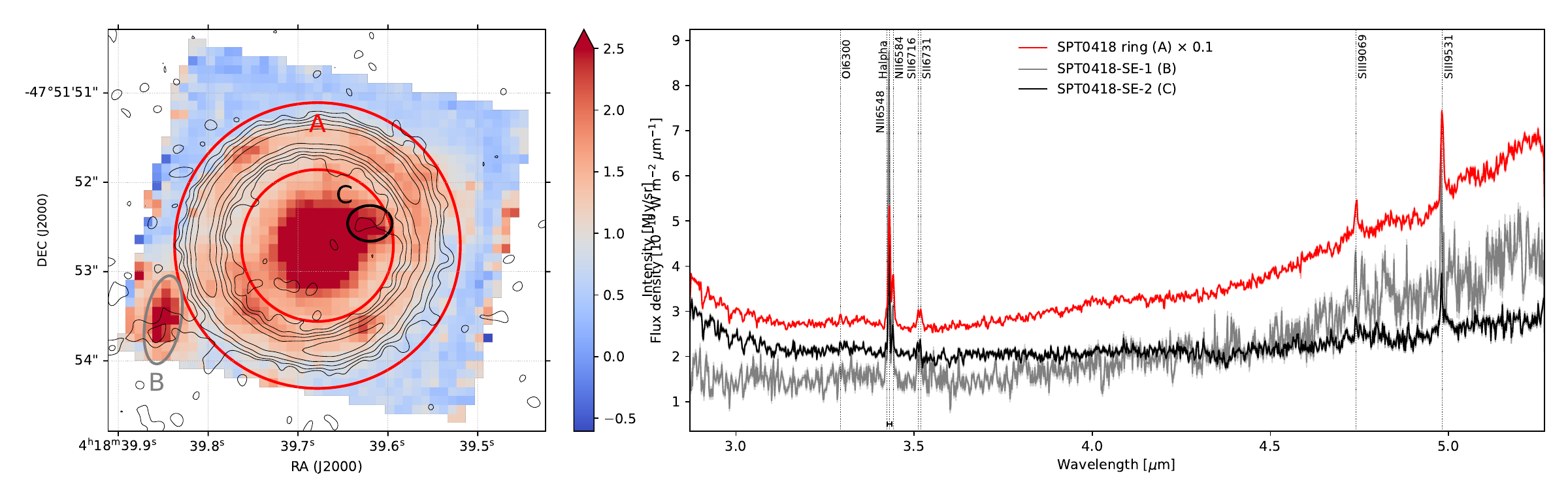}
	\caption{Left: \ha pseudo-narrowband image of the SPT0418 system, averaged over the channels including the \ha emission in the original spectral cube. The strongly lensed ring and the two newly discovered sources (SE-1 and SE-2) are highlighted by red annulus, magenta and black ellipses, marked as ``A'', ``B'' and ``C'' respectively. The lensing galaxy is shown as the central bright source. The 835 \micron\, continuum is plotted as the thin black contour, with the levels $2, 4, 8, 16, 32 \times \sigma$ where $\sigma = 56.7 \ \mathrm{\mu Jy\ beam^{-1}}$. Right: the spectra of the three sources integrated over the regions highlighted in the left panel, using the same colour scheme. The spectrum for the ring is scaled by a factor of 0.1 for clarity. The small black bar below the \ha line marks the wavelength coverage of the pseudo-narrowband image. The potentially detected lines are marked by vertical dotted lines.}
	\label{f:spec}
\end{figure*}

We create an image (Fig.~\ref{f:spec} left) by collapsing all emission in spectral channels covering the \ha line in the original NIRSpec spectral cube.
This pseudo-narrowband image reveals the presence of two sources in addition to the strongly lensed ring and the lensing elliptical galaxy.
The brighter source at the south east corner (``B'' in Fig.~\ref{f:spec}) is referred to as SPT0418-SE-1, and the source between the lensing galaxy and ring (``C'') is denoted SPT0418-SE-2.

The spectra of the newly-discovered sources as well as the strongly-lensed ring (Fig.~\ref{f:spec} right) show strong detection of the \ha line, [N II]$\lambda6548, 6584$ doublet, [S II]$\lambda6716, 6731$ doublet, [S III]$\lambda9069, 9531$ lines, and a tentative detection of the [O I]$\lambda6300$ line.
These strong emission lines enable us to identify the association of the discovered sources in redshift space.

Both SE-1 and SE-2 also appear in NIRCam and MIRI images from 1-10\,\micron. 
Based on this knowledge, we also performed forced photometry on the ALMA data and found that SE-1 appears in the 835 \micron\ continuum map with a $\sim 4 \sigma$ peak, while SE-2 is associated with a tentative $2 \sigma$ peak (Fig.~\ref{f:spec} left contour). 
Both of them are also detected in the \cii 158 \micron\, line adding to the spectroscopic confirmation of the redshift of the newly discovered sources.

\subsection{Observational Properties}
\label{sec:properties}

\begin{figure*}
	\centering
	\includegraphics[width=1\textwidth]{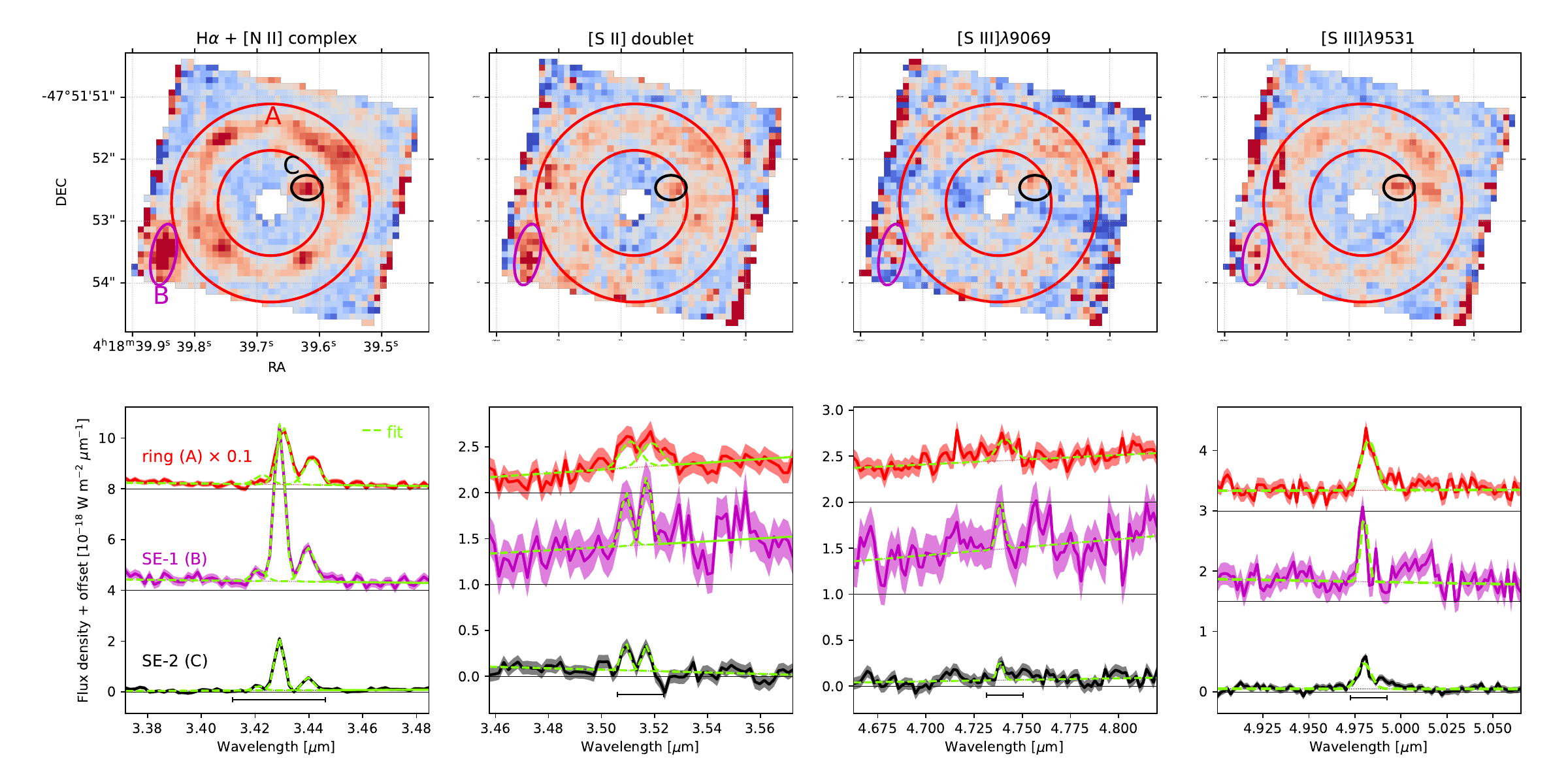}
	\caption{The moment 0 map and spectrum snippets of the strong emission lines. For each strong line(s), the lens-subtracted continuum subtracted moment 0 map is displayed in the first row, with the same marked regions as in Fig.~\ref{f:spec}. The spectrum snippets are shown in the second row, integrated over the three regions. The spectra are offset vertically to aid the vision, with the zero baseline indicated by a thin black line. The yellow dashed line indicates the Gaussian fit to the line or line associations plus the linear fit to the local continuum. The plotting style and colour of the spectrum are the same as Fig.~\ref{f:spec}.}
	\label{f:stamps}
\end{figure*}

\begin{figure*}
	\centering
	\includegraphics[width=1\textwidth]{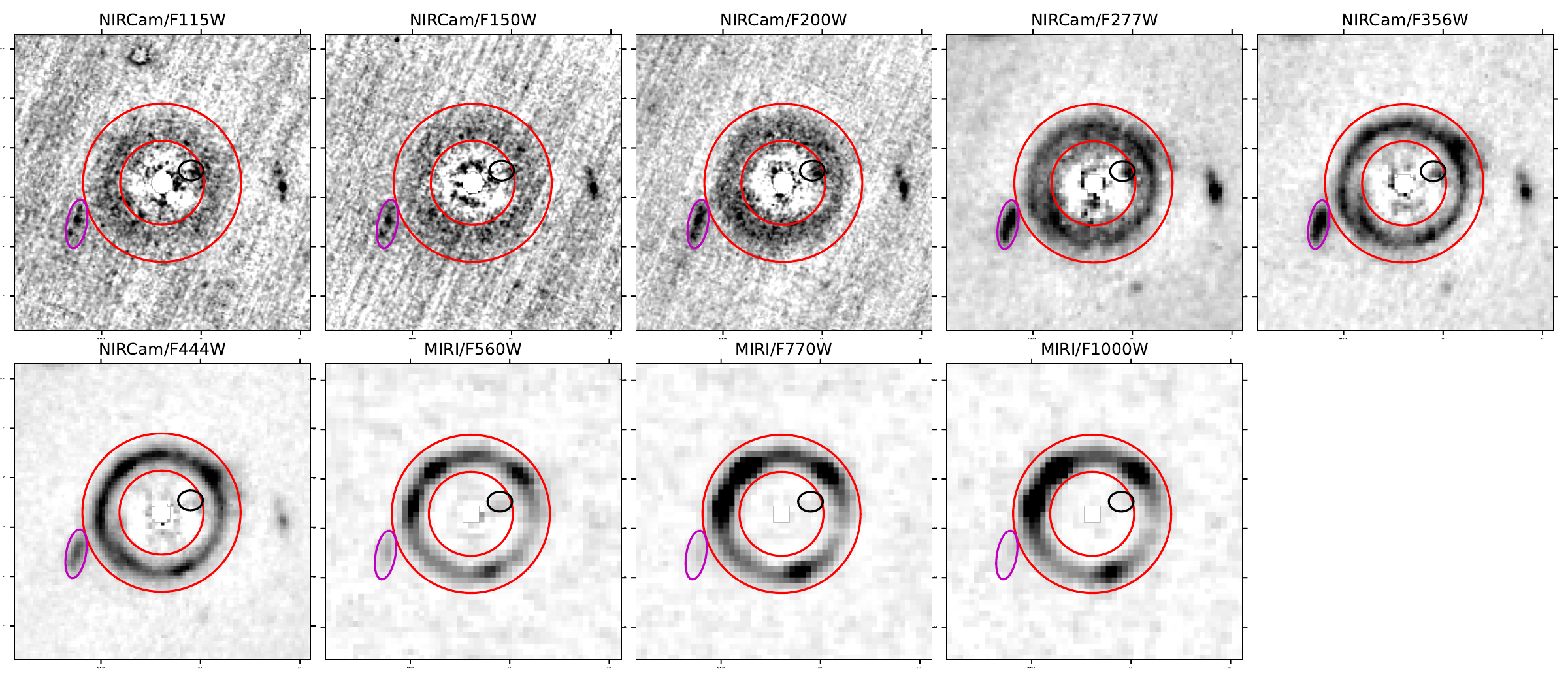}
	\caption{Atlas of lens-subtracted NIRCam and MIRI images in the wavelength range $\lambda_\mathrm{obs}=1.15 - 10\ \micron$, corresponding to $\lambda_\mathrm{rest}=0.22 - 1.9\ \micron$. The filter is shown as the title of each stamp. The images are oriented and cropped to fit the image stamps in the previous images, so are the highlighted regions and colour scheme. The colour map is adjusted manually to increase the contrasts of the weak sources.}
	\label{f:atlas}
\end{figure*}

We show the moment-0 maps and spectra for the strong lines in Fig.~\ref{f:stamps}.
Different from the pseudo-narrowband image, the moment-0 maps are created using the lens-subtracted data, and the continuum is removed using a linear fit to the local spectrum. 
The spectra are extracted from the lens-subtracted cube, and the emission line is modeled and fitted as a Gaussian profile along with a first order polynomial fit on the local continuum, in order to to measure the line flux and width.
In the fitting procedure, the redshift is fixed relative to that of \ha, and the doublets are fitted as a whole using known emissivity ratio (for [N II]) and the same line width (for the [N II] and [S II] doublets).

The SE-1 and 2 sources show up strongly in all the moment 0 maps.
Remarkably, in the \ha + {[N II]} map, SE-1 has an even higher surface brightness than the ring.

In all the spectra (Fig.~\ref{f:stamps}, especially \ha), the line peaks of both SE-1 and SE-2 lie consistently at -175 km/s relative to the ring, and the fitted line widths are slightly narrower (NIRSpec spectral resolution $\sim$ 110 km/s).
The line center wavelengths and the widths are consistent between SE-1 and SE-2 with a SE-1 to SE-2 flux ratio about 3:1.
One caveat is that the fitted line widths for detections with SNR$<$10 are not well constrained and should not be compared quantitatively.
We also caution the absolute wavelength calibration of NIRSpec data is not optimal, as the ALMA \cii line is systematically offset at a lower redshift for both the ring and the newly-discovered sources (see table~\ref{t:flux}).

The 1 to 10 \micron\, lens-subtracted images are shown in Fig.~\ref{f:atlas}.
We note that at wavelengths shorter than $\lambda < 2\ \micron$, SE-1 appears to have a bright nucleus accompanied by two faint patches on both north and south sides.
The NIRCam 2.8 and 3.6 \micron\, images are strongly affected by the [O III]$\lambda$5007 and \ha emission lines respectively, SE-1 appear more extended, and the surface brightness contrast to the ring are higher compared to other wavelengths.
The primarily lensed galaxy (ring) also shows extended morphology that agrees with the line maps in Fig.\ref{f:stamps}.
Only the images up to 10 \micron\, are plotted as the companion sources are barely detected at longer wavelengths.

The dark hexagon shape present at the center of the short wavelength images is the result of the compact continuum emission from the lensing galaxy (Sec.~\ref{sec:jwst}).
This PSF residual makes the identification and photometry challenging for SE-2. 
Further analysis involving subtraction of the lensing galaxy would require a core model convolved with JWST PSF and is a subject of future publication.

\begin{table*}
\centering
\caption{Properties of the newly discovered source. The ``SE'' column is the combined properties of both SE-1 and SE-2, as we speculate that they are the primary and counter lensed image of the same galaxy. For non-detection in SE-2 we report the $3\sigma$ upper limit, and scale the SE-1 value up by a factor of 1.3 to estimate the value for SE.}
\include{spt0418_se_spec_flux}
\label{t:flux}
\end{table*}

The basic source properties like location, line flux and width, and photometry from NIRCam and MIRI images are summarized in Table~\ref{t:flux}.
Line fluxes and full width half maximum (FWHM) are derived using a Gaussian fit.

\section{Discussion}
\label{sec:discussion}

\subsection{Nature and Morphology}
\label{sec:nature}

Based on the redshift, line profile, and spatial separation between the newly identified source and the ring, it is evident that both SE-1 and SE-2 are physically different sources than the primarily lensed galaxy of the Einstein ring. 
We further argue that SE-1 and SE-2 are in fact two lensed images of the same galaxy: both sources have the same redshift and line width in the strong ($> 10 \sigma$) line detections within the allowance of measurement error; the line fluxes and continuum consistently show a SE-1 to SE-2 ratio roughly 3:1; and the location of SE-1 and SE-2 measured with respect to the ring is consistent with them being lensed images of a source separated from the primarily lensed galaxy (ring) by about 0\farcs7, in the south east direction.

Therefore, we suggest that SE-1 and SE-2 are the primary and counter images of a companion galaxy (hereafter referred to as companion or SE) lensed by the same foreground galaxy as the ring (hereafter as the host galaxy).
As gravitational lensing non-linearly distorts the actual physical geometry, based on the separation of SE-1 and SE-2 in the image plane (0\farcs7), we estimate that the companion must also be within a projected distance of 5 kpc from the host in the source plane. 
From the lensing models for the ring presented in \citet{spilker16,rizzo20}, we know that this galaxy-galaxy lensing system has a relatively simple gravitational potential and line of sight alignment (small impact parameter).
We can apply basic gravitational lensing physics \citep{blandford92}, to posit that the secondary image (SE-2) has a magnification factor $\mu \sim 1$. 
Furthermore, based on the linear scale of the elongated shape, we estimate that the magnification of SE-1 is at least 8 times smaller than the ring, and estimate an upper limit for the total magnification of SE is $\mu_{\rm max} < 6$. 
Detailed gravitaional lens modelling involving reconstruction of the sources and their intrinsic arrangement in the source plane is beyond the scope of this letter. 

The morphology of the companion is intriguing. 
At short wavelengths, the galaxy displays a nuclear feature, which we will argue in Sec.~\ref{sec:rad} to be a starburst nucleus.
The morphology of gas and the old stellar population by $\lambda_{rest}=0.8-2\ \micron$ images, are more extended than the dust traced by ALMA continuum and young stellar population by $\lambda_{rest} \leq 0.4\ \micron$ images.
This dichotomy appears in both the SE-1\&2 and the ring, and is consistent with the recent findings that \cii is often more extended than both UV and dust continuum \citep{carniani18,fujimoto19,ginolfi20,fudamoto22}. 
In addition, there is a dust lane-like gap at the southeast and northwest of the ring in the \ha map, 2.8 to 4.4 \micron\ images (the broken connection of the ring at the lower left and upper right part), suggesting its association with gas emission.
We suspect the gaps are regions of very high dust extinction or the line emission is suppressed.
In addition to the newly discovered companion, there are even more arc-like weak features in NIRCam images (see Fig.~\ref{f:atlas} for example) beyond the NIRSpec field of view. 
Future deeper and wider spectroscopic observations may decipher this seemingly crowded field.

\subsection{Dust Obscured Star Formation Rate}
\label{sec:sfr}

\begin{figure}
	\centering
	\includegraphics[width=0.5\textwidth]{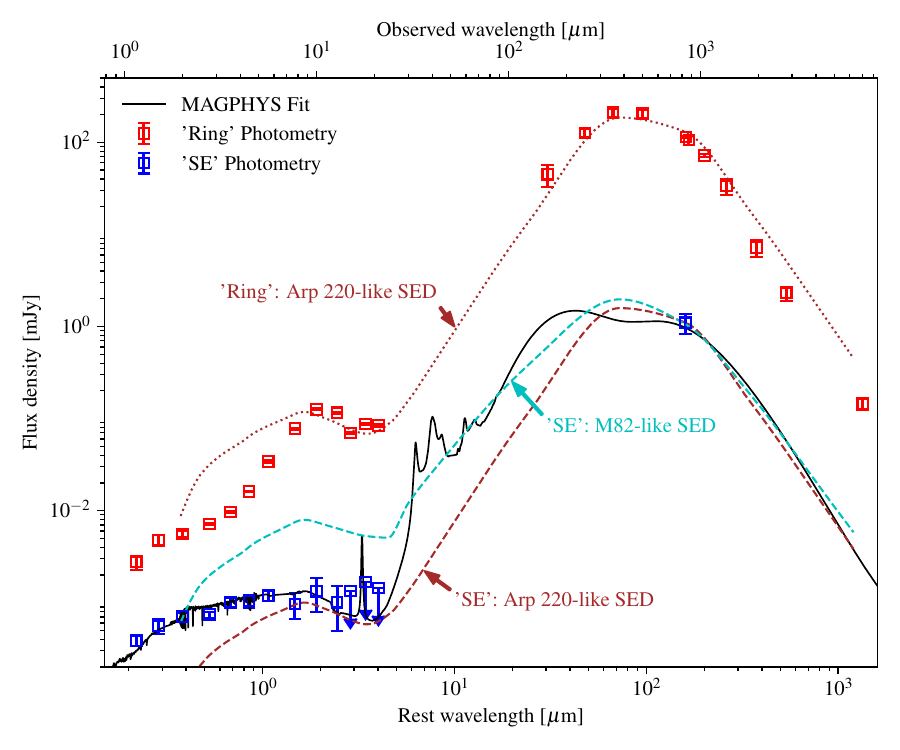}
	\caption{SED of \spt system (red square) and the companion SE (blue square), along with the Arp-220 (red dashed and dotted) and M82 (cyan dashed) SED templates \citep{polletta07} normalized by the dust continuum. The figure is adapted from \citep{vishwas19}.}
	\label{f:sed}
\end{figure}

We estimate the star formation rate of the companion galaxy in three ways: SFR$_\mathrm{H\alpha}$ using the formula in \citet{murphy09}, SFR$_\mathrm{[C\ II]}$ using \cii-SFR relation in \citet{camus15}, SFR$_\mathrm{FIR}$ using the scaling relation given by \citet{kennicutt98}.
As we have a single point on the dust SED for the companion, we estimate the FIR luminosity assuming a $\beta=1.8$ and the commonly used 850 \micron\ relation \citep{genzl10,carilli13}. 
The estimates of SFR are not corrected for lensing and are thus labelled as $\mu$SFR.
The results are summarized in Table~\ref{t:flux}.

We found that while the SFR$_\mathrm{[C\ II]}$ agrees with SFR$_\mathrm{FIR}$ within errors due to the limited fidelity afforded by using a single photometric point, it is a factor of $>$ 2 times higher than SFR$_\mathrm{H\alpha}$. 
By attributing all the difference between SFR$_\mathrm{[C\ II]}$ and SFR$_\mathrm{H\alpha}$ to dust extinction, we can obtain an obscuration fraction of star formation $> 55\%$ for SE, which is at the higher end for the galaxies with simialr stellar mass at z=4$\sim$ 5 \citep{faisst22}.
This high obscuration fraction suggests that the companion galaxy is also dusty in nature.
Although the in-flight calibration of JWST instruments is still ongoing, and the absolute flux calibration of NIRSpec may be updated in the future, we do not anticipate any order of magnitude correction in calibration
And as a consistency check we compared the continuum flux density measured in the NIRSpec spectrum and find it matches well with the photometric measurements on the better calibrated NIRCam and MIRI images.

Using the upper limit of the magnification for the companion galaxy SE, and a reconciled $\mu$SFR = 100 M$_\sun$/yr, we can obtain a lower limit of the intrinsic SFR of 17 M$_\sun$/yr.
This intrinsic SFR is about 20 times less than the host galaxy \citep{debreuck19}.
This securely places SE in the range of typical star-forming galaxies at redshift 4-5 \citep{lefevre20}.

The high dust obscuration is also apparent in the spectral energy distribution (SED) in Fig.~\ref{f:sed}. 
A large infrared-to-optical excess (IRX) is present for the host galaxy at a level comparable to the local ULIRG Arp-220 with IRX=3.4 \citep{howell10}.
SE shows slightly less obscuration in SED than the host, while still being more extreme than the typical local starburst galaxy M82.
This difference in obscuration might be responsible for the higher \ha surface brightness in SE-1 than the ring.

SED fitting carried out with high-z extension to MAGPHYS \citep{dacunha08,dacunha15} shows SE has a stellar mass $\mu\mathrm{M_\star} = 7.0^{+3.9}_{-2.3}\times 10^9~\mathrm{M_\sun}$, placing it at the lower mass end for star-forming galaxies at this redshift \citep{rizzo20}, but above the star forming main sequence \citep{martin15}.
The SED based $\mu\mathrm{SFR} = 58^{+15}_{-23}~\mathrm{M_\sun/yr}$ and $\mu \mathrm{L_{IR}} = 7.4^{+7.0}_{-3.2} \times 10^{11}~\mathrm{L_\sun}$ agree with our estimates for SE using other methods, and clearly highlight the $\sim$2$\times$ uncertainty for estimating infrared luminosity using a single photometric data point.
The SED fitting also yields a depth $\tau_V = 1.49^{+1.52}_{-0.5}$ and dust mass $\mu\mathrm{M_{d}} = 1.5^{+1.5}_{-0.8} \times 10^8~\mathrm{M_\sun}$, suggesting a significant obscuration by dust content accounting for up to 2\% of the stellar mass in this galaxy. 
Although we still caution the poor constraints provided by the single data point in the rest-frame FIR band.

Most (U)LIRGs in the local Universe are in the process of merging, which likely triggers the starburst activity in these systems \citep{sanders96,armus09}.
It is thus natural to speculate \spt as an analog of a ULIRG triggered by a merger at z$\sim$4.2. 
The small size of the companion and the distortion by gravitational lensing, lead to some uncertainty to study the relation between the host and the companion galaxies in 3D and whether it can be classified as a major or a minor merger.
The physical proximity, both in projected distance and redshift space, supports the merger scenario and also raises questions with respect to the dynamical interaction timescale. 
These could potentially be in contradiction to the observed low velocity dispersion or the "dynamically cold" disk scenario suggested by \citet{rizzo20}, which will be disrupted by either tidal interactions that are commonly seen in the local (U)LIRGs, or the cold accretion expected for high redshift minor mergers.

\subsection{Radiation field}
\label{sec:rad}

We characterised the radiation field using the Nitrogen and Sulfur lines detected in the NIRSpec spectrum. 
The {[N II]}$\lambda$6584/\ha ratio $\sim$ 0.25 places SE in the star forming locus in BPT diagram with relatively soft radiation \citep{bpt,kewley19}.
This leads to the argument that the compact nuclear feature present in the $\lambda_\mathrm{rest} \leq 5000\ \mathrm{\AA}$ images is more likely a starburst nucleus instead of an AGN. 
While we can not rule out a low-luminosity AGN being present, the star-formation scenario is further supported by the lack of strong rest-frame UV radiation, dimming in mid-infrared bands, and the high SFR inferred in this galaxy.

The $S32$ index of $\log$ {[S III]}/{[S II]} is primarily a measure of the ionization parameter U.
The shorter wavelength {[S II]} line is more extincted, so we can only put an upper limit on $S32$ ratio to be $\lesssim$0.25.
We then estimated an upper limit of $\log U \lesssim -3$ using the diagnostics given in \citep{sanders20}.
This is within the range measured in the local Universe, suggesting similar conditions in the ionized gas of the star forming region \citet{kewley19}.
The small value of U also predicts relatively weak {[O III]}$\lambda$4959,5007 lines \citep{strom18}, as is also suggest by its allowed position on BPT diagram.

\subsection{Metallicity}
\label{sec:metallicity}

We estimated the chemical abundance in SE using three methods.
The absolute abundance O/H is calculated using two strong line methods: the $N2$ index with the third-order polynomial fit in \citetalias{pettini04}; and the $S_{23}$ index in \citet{diaz00}.
We also used the $N2S2$ index \citep{viironen07,perez09} to measure the Nitrogen-to-Oxygen abundance ratio N/O.
The results are recorded in Table~\ref{t:flux}.

We found SE to be chemically mature with O/H $\sim$ 0.6 times the solar abundance and N/O $\sim$ 1.2 times solar, which are surprisingly high for a galaxy only 1.46 billion years after the Big Bang.
We argue that the metallicity estimates are robust because (1) only line ratios are used and they are not affected by the uncertainties in the absolute calibration of JWST; (2) the emission lines are close in wavelength, so the line ratio is not subject to differential extinction, with the only exception being the {[S III]} lines; (3) the three methods use a suite of lines and all reach a consistent result of near-solar metallicity; it would require at least two of the line-flux values to change in order to attain some other consistent results.
Systematic uncertainty in these empirical optical diagnostics are typically 0.3 dex \citep{perez09}, but it is not included in the error values reported in Table~\ref{t:flux} for clarity. 
However, the systematic error is reduced when combining and comparing different methods.
Besides, the SED fitting of SE using MAGPHYS also yields a reasonably consistent metallicity $\sim 0.78\ Z_\sun$.

The near-solar abundance and the elevated N/O ratio, suggests this galaxy hosts highly enriched interstellar medium. 
It is consistent with the large dust content and relatively soft radiation field in this galaxy. 
But drawing a comparison with the metallicity of star-forming galaxies at redshift 1 to 3 from ground based observations \citep[c.f.][]{steidel14,kashino17,strom22}, both O/H and N/O of SE are at the very high end of the distribution, yet they are at a much earlier epoch in the Universe. 
This suggests a rare and unusual chemical evolution history for SE.

It is even more striking that an identical analysis on the host galaxy found the metallicity Z $\sim$ 1.6 Z$_\sun$, and (N/O) $\sim$ 3 $\times$ (N/O)$_\sun$.
This metallicity is even higher than $\mathrm{Z}\sim\mathrm{Z_\sun}$ derived in \citet{debreuck19}.
Because the gas phase elemental abundance builds up through nuclear synthesis and stellar feedback, and the growth of N/O is slower through the prolonged secondary production of Nitrogen, O/H and N/O abundances strongly constrains the age and star forming efficiency (SFE) of a galaxy. 
Comparing with the chemical evolution model in Fig. 5 of \citet{vincenzo16}, the stellar population in SE is inferred to have an age of at least 800 Myr with a preferentially short depletion time scale $\tau \leq 0.67 \ \mathrm{Gyr}$, while the super-solar metallicity of the host, especially the high N/O, requires the galaxy formation to start shortly after the Big Bang, and being sustained at a high SFE with $\tau \leq 0.3 \ \mathrm{Gyr}$. 

The high SFE inferred from the metallicity supports a decreasing depletion time towards higher redshift found by other studies \citep{saintonge13,tacconi20}. 
The younger age of the companion suggested by the lower metallicity is also consistent with the lack of $\lambda_\mathrm{rest} \sim 2\ \micron$ peak in the SED corresponding to the old stellar population, which shows up strongly for the more evolved host galaxy. 
On the other hand, the inferred early onset of galaxy formation, clustering of galaxies, and the potential crowded field indicates the system might trace a massive dark matter halo that enables structure formation at a very early time in the Universe.

Although the FIR fine-structure line observations are not resolved for SE, we can still compare the FIR diagnostic of the whole system. 
In addition to \cii, the system is also detected in [O\,{\sc iii}]\ 88 \micron\ [N\,{\sc ii}]\ 122 \micron\ and [N\,{\sc ii}]\ 205 \micron\ lines \citep{debreuck19,cunningham20}.
The [N\,{\sc ii}]122/[O\,{\sc iii}]88 flux ratio, which can be a rough proxy for N/O, is only 0.05. 
In the local universe, such a small line ratio is more commonly seen in dwarf galaxies in consequence of both hard stellar radiation fields and low N/O \citep{cormier15}, rather than (U)LIRGs \citep{diaz17} that the dusty and metal rich \spt system resembles. 
The weak [N\,{\sc ii}]\ 122 and 205 \micron\ lines are also in contrast to the relatively bright [N II]$\lambda$6584 line.

One possible solution is the applicability of the optical strong line diagnostics in the early Universe. 
\citet{strom18} found a slightly modified relation for estimating the elemental abundances using optical spectral lines at redshift 2 -- 3 compared to those from local galaxies.
Using the revised $N2$ and $N2S2$ calibration in \citet{strom18}, the N/O of both the host and the companion decrease by about 0.25 dex, corresponding to 1.6 and 1 times the solar abundance respectively. 
However, we do not find a significant change in the estimate for {O/H}. 
Though still high, the new values of N/O need a much less evolved stellar population, and they are closer to the $\mathrm{\log(N/O)}\sim-1.1$ estimated from the empirical [O\,{\sc iii}]88/[N\,{\sc ii}]122-to-N/O relation in the local Universe (Peng in prep.).
We leave more detailed analysis on chemical abundance of the ring to future publications.

\section{Summary and Future Work}

In this study, we report the discovery of a dusty star-forming companion galaxy \spt-SE in the \spt system. 
The companion galaxy is also gravitationally lensed, which results into the two images SE-1 and SE-2. 
By combining JWST and ALMA observations, we find a high SFR of the companion galaxy with near unity dust obscuration fraction. 
Using strong line indices we find near-solar and super-solar metallicity for the companion and the host galaxy. 

This work highlights the capability of JWST for discovering fainter and lower SFR galaxies in the early Universe. 
It also shows the value of joint study of both optical and submillimeter observations, as 55\% SFR of the companion galaxy is obscured by dust. 
JWST for the first time enables us to study the physical conditions of the hot gas through strong optical lines at z$>$4. 
This work illustrates some exciting discoveries and results that are attained with these lines. 
This spectroscopic study of a $z>4$ galaxy opens up many questions including: the spatial arrangement and stellar/gas/metallicity distribution of the companion; the merging hypothesis of \spt; the dark matter halo of the system; the over-density of this potentially crowded field; reconciling the relatively high chemical abundances with the short formation time and the moderate stellar mass for the whole system; and interpreting the small {[N\,\sc{ii]}} 122 and 205 \micron\ luminosities in the context of either a soft radiation field and/or a high N/O.

Our work suggests \spt is a good example of the very early mass buildup and structure formation based on the solar-like metallicity at a cosmic age of 1.4 Gyr and hints of clustering .
The high lensing factor, rich observational data and the many intriguing questions on its formation history make it an ideal target for future observations and in-depth study. 
We therefore call for NIRSpec observations covering the {[O III]}$\lambda$4959,5007 doublets and [O II]$\lambda$3727 lines, an \ha narrowband search in the \spt field, as well as deep resolved ALMA {[N\,\sc{ii}]} and {[O\,\sc{iii}]} observations.

\begin{acknowledgments}

The authors thank the anonymous referee for the helpful comments and suggestions.
This work was supported by the National Science Foundation through the following grants: NRAO SOS 1519126, AST-1716229, CAREER-1847892 and AST-2009767, and NASA/USRA SOFIA grant 09-0185.
This work is based on observations made with the NASA/ESA/CSA James Webb Space Telescope. The data were obtained from the Mikulski Archive for Space Telescopes at the Space Telescope Science Institute, which is operated by the Association of Universities for Research in Astronomy, Inc., under NASA contract NAS 5-03127 for JWST. These observations are associated with program 1355. 
The authors acknowledge the TEMPLATES team led by Jane Rigby for developing their observing program with a zero-exclusive-access period.
The National Radio Astronomy Observatory is a facility of the National Science
 Foundation operated under cooperative agreement by Associated Universities, Inc.
This paper makes use of the following ALMA data: ADS/JAO.ALMA\#2016.1.01374.S ALMA is a partnership of ESO (representing its member states), NSF (USA) and NINS (Japan), together with NRC (Canada), MOST and ASIAA (Taiwan), and KASI (Republic of Korea), in cooperation with the Republic of Chile. The Joint ALMA Observatory is operated by ESO, AUI/NRAO and NAOJ. The National Radio Astronomy Observatory is a facility of the National Science Foundation operated under cooperative agreement by Associated Universities, Inc.

\end{acknowledgments}

\bibliography{main}{}
\bibliographystyle{aasjournal}

\end{document}

%% file: spt0418_se_spec_flux.tex
\begin{tabular}{cccccc}
\hline
\hline
   & \multicolumn{2}{c}{SPT0418-SE-1} & \multicolumn{2}{c}{SPT0418-SE-2} & SPT0418-SE\\
\hline
   R.A. (J2000) & \multicolumn{2}{c}{04:18:39.85(1)} & \multicolumn{2}{c}{04:18:39.62(1)} & \\
   Decl. (2000) & \multicolumn{2}{c}{-47:51:53.54(5)} & \multicolumn{2}{c}{-47:51:52.46(5)} & \\
   z$_\mathrm{H\alpha}$ & \multicolumn{2}{c}{4.22507(1)} & \multicolumn{2}{c}{4.22506(1)} & 4.22507(1)\\
   z$_\mathrm{[C\ II]}$ & \multicolumn{2}{c}{4.223} & \multicolumn{2}{c}{4.223} & 4.223\\
\hline
   & \multicolumn{5}{c}{Spectroscopic properties} \\ 
  line & Flux & FWHM & Flux & FWHM & Flux\\
   & $10^{-21}$ W m$^{-2}$ & km s$^{-1}$ & $10^{-21}$ W m$^{-2}$ & km s$^{-1}$ & $10^{-21}$ W m$^{-2}$\\
\hline
  H$\alpha$ & 29.40 $\pm$ 0.66 & 390 $\pm$ 11 & 8.94 $\pm$ 0.23 & 358 $\pm$ 13 & 38.34 $\pm$ 0.70\\
  {[N II]}$\lambda$6548 & 2.78 $\pm$ 0.75 & 507 $\pm$ 59 & 0.84 $\pm$ 0.25 & 408 $\pm$ 51 & 3.62 $\pm$ 0.79 \\
  {[N II]}$\lambda$6584 & 8.27 $\pm$ 0.75 & 507 $\pm$ 59 & 2.50 $\pm$ 0.24 & 408 $\pm$ 51 & 10.77 $\pm$ 0.79 \\
  {[S II]}$\lambda$6716 & 2.78 $\pm$ 0.80 & 380 $\pm$ 76 & 1.17 $\pm$ 0.25 & 312 $\pm$ 61 & 3.95 $\pm$ 0.84 \\
  {[S II]}$\lambda$6731 & 3.52 $\pm$ 0.80 & 380 $\pm$ 76 & 1.10 $\pm$ 0.25 & 312 $\pm$ 61 & 4.62 $\pm$ 0.84 \\
  {[S III]}$\lambda$9069 & 2.41 $\pm$ 0.80 & 270 $\pm$ 114 & 0.85 $\pm$ 0.23 & 240 $\pm$ 87 & 3.26 $\pm$ 0.83 \\
  {[S III]}$\lambda$9531 & 5.45 $\pm$ 0.72 & 282 $\pm$ 49 & 3.99 $\pm$ 0.54 & 527 $\pm$ 72 & 9.44 $\pm$ 0.90 \\
  {[C\,\sc{ii}]}158\micron & 55.9 $\pm$ 8.5 & 200 $\pm$ 70 & 17.7 $\pm$ 3.6 & 200 $\pm$ 90 & 73.6 $\pm$ 9.2 \\
   & ($4.6 \pm 0.7\ \mathrm{Jy\ km/s}$) & & ($1.46 \pm 0.30 \ \mathrm{Jy\ km/s}$) & & ($6.06 \pm 0.76 \ \mathrm{Jy\ km/s}$) \\
\hline
   & \multicolumn{5}{c}{Photometric properties} \\ 
  filter & \multicolumn{2}{c}{S$_{\nu}$}  & \multicolumn{2}{c}{S$_{\nu}$} & S$_{\nu}$ \\
   & \multicolumn{2}{c}{${\mu}$Jy}  & \multicolumn{2}{c}{${\mu}$Jy} & ${\mu}$Jy \\
\hline
  NIRCam/F115w & \multicolumn{2}{c}{0.272 $\pm$ 0.019} & \multicolumn{2}{c}{0.114 $\pm$ 0.040} & 0.386 $\pm$ 0.044\\
  NIRCam/F150w & \multicolumn{2}{c}{0.429 $\pm$ 0.026} & \multicolumn{2}{c}{\textless 0.286} & 0.558 $\pm$ 0.099\\
  NIRCam/F200w & \multicolumn{2}{c}{0.547 $\pm$ 0.020} & \multicolumn{2}{c}{0.163 $\pm$ 0.050} & 0.710 $\pm$ 0.054\\
  NIRCam/F277w & \multicolumn{2}{c}{0.608 $\pm$ 0.020} & \multicolumn{2}{c}{0.134 $\pm$ 0.052} & 0.742 $\pm$ 0.056\\
  NIRCam/F356w & \multicolumn{2}{c}{0.845 $\pm$ 0.022} & \multicolumn{2}{c}{\textless 0.348} & 1.01 $\pm$ 0.118\\
  NIRCam/F444w & \multicolumn{2}{c}{0.808 $\pm$ 0.021} & \multicolumn{2}{c}{\textless 0.508} & 1.05 $\pm$ 0.171 \\
  MIRI/F560w & \multicolumn{2}{c}{0.830 $\pm$ 0.068} & \multicolumn{2}{c}{0.358 $\pm$ 0.141} & 1.19 $\pm$ 0.156 \\
  MIRI/F770w & \multicolumn{2}{c}{0.740 $\pm$ 0.106} & \multicolumn{2}{c}{0.227 $\pm$ 0.286} & 0.967 $\pm$ 0.305 \\
  MIRI/F1000w & \multicolumn{2}{c}{1.02 $\pm$ 0.211} & \multicolumn{2}{c}{\textless 1.48} & 1.33 $\pm$ 0.536 \\
  MIRI/F1280w & \multicolumn{2}{c}{0.776 $\pm$ 0.291} & \multicolumn{2}{c}{\textless 1.28} & 1.01 $\pm$ 0.516 \\
  MIRI/F1500w & \multicolumn{2}{c}{\textless 0.963} & \multicolumn{2}{c}{\textless 0.932} & \textless 1.34 \\
  MIRI/F1800w & \multicolumn{2}{c}{\textless 1.08} & \multicolumn{2}{c}{\textless 1.28} & \textless 1.67 \\
  MIRI/F2100w & \multicolumn{2}{c}{\textless 0.918} & \multicolumn{2}{c}{\textless 1.11} & \textless 1.44 \\
  ALMA/835$\mu$m & \multicolumn{2}{c}{733 $\pm$ 227} & \multicolumn{2}{c}{364 $\pm$ 150} & 1100 $\pm$ 272 \\
\hline
   & \multicolumn{5}{c}{Derived properties} \\ 
  $\mu$L$_\mathrm{H\alpha}$ {[$10^8$ L$_\sun$]} & \multicolumn{2}{c}{13.9 $\pm$ 0.31} & \multicolumn{2}{c}{4.24 $\pm$ 0.11} & 18.2 $\pm$ 0.33\\
  $\mu$L$_\mathrm{[C\ II]}$ {[$10^8$ L$_\sun$]} & \multicolumn{2}{c}{26 $\pm$ 4.0} & \multicolumn{2}{c}{8.41 $\pm$ 1.72} & 34.4 $\pm$ 4.4\\
  $\mu$L$_\mathrm{FIR}$ {[$10^{11}$ L$_\sun$]} & \multicolumn{2}{c}{$9.1^{+6.7}_{-3.6}$} & \multicolumn{2}{c}{$2.9^{+2.7}_{-2.0}$} & $12.0^{+8.7}_{-4.5}$ \\
  $\mu$SFR$_\mathrm{H\alpha}$ {[M$_\sun$/yr]} & \multicolumn{2}{c}{29$\pm$ 0.6} & \multicolumn{2}{c}{ 8.8 $\pm$ 0.2} & 38 $\pm$ 0.7 \\
  $\mu$SFR$_\mathrm{[C\ II]}$ {[M$_\sun$/yr]} & \multicolumn{2}{c}{65 $\pm$ 10} & \multicolumn{2}{c}{20 $\pm$ 4.2} & 87 $\pm$ 11 \\
  $\mu$SFR$_\mathrm{FIR}$ {[M$_\sun$/yr]} & \multicolumn{2}{c}{$118^{+87}_{-47}$} & \multicolumn{2}{c}{$38^{+35}_{-26}$} & $156^{+113}_{-58}$ \\
  $\log(\mathrm{O/H)}_{N2}$ & \multicolumn{2}{c}{-3.42 $\pm$ 0.10} & \multicolumn{2}{c}{-3.42 $\pm$ 0.11} & -3.42 $\pm$ 0.08 \\
  $\log(\mathrm{O/H})_{S23}$ & \multicolumn{2}{c}{-3.52 $\pm$ 0.34} & \multicolumn{2}{c}{-3.18 $\pm$ 0.29} & -3.42 $\pm$ 0.26 \\
  $\log(\mathrm{N/O})_{N2S2}$ & \multicolumn{2}{c}{-0.71 $\pm$ 0.21} & \multicolumn{2}{c}{-0.80 $\pm$ 0.18} & -0.73 $\pm$ 0.16 \\
\hline
\end{tabular}